# Revisiting the elastic solution for an inner-pressured functionally graded thick-walled tube within a uniform magnetic field

Libiao Xin [a, b, *], Yanbin Li [c], Dongmei Pan [d], Guansuo Dui [e], Chengjian Ju [e].

[a] Shanxi Key Laboratory of Material Strength & Structural Impact, College of Mechanics, Taiyuan University of Technology, Taiyuan, 030024, China

[b] National Demonstration Center for Experimental Mechanics Education, College of Mechanics, Taiyuan university of Technology, Taiyuan 030024, China

[c] Applied Mechanics of Materials Laboratory, Department of Mechanical Engineering, Temple University, Philadelphia, Pennsylvania 19122, USA

[d] Department of Civil and Environmental Engineering, University of Houston, Houston, Texas 77204, USA

[e] Institute of Mechanics, Beijing Jiaotong University, Beijing, 100044, China

**Abstract**

In this paper, the mechanical responses of a thick-walled functionally graded hollow cylinder subjected to uniform magnetic field and inner-pressurized loads are studied. Rather than directly assuming the material constants into some certain function forms as displayed in pre-researches, we firstly give the volume fractions of different constituents of the FGM cylinder and then determine the expressions of material constants. By the use of the Voigt method the corresponding analytical solutions of displacements in radical direction, the strain and stress components and the perturbation magnetic field vector are derived by the following. In numerical part,

---

\* Corresponding author

E-mail address: xinlibiao@tyut.edu.cn

the influences of volume fraction on displacement, stain and stress components and the magnetic perturbation filed vector are investigated; indicated by the results, we can conclude that the Poisson's ratio has a significant effect on the FGM cylinder's mechanical behaviors. Moreover, by some appropriate choices of the material constants it can be found that the obtained results in this paper can reduce to some special cases given in previous literatures.

**1 Introduction**

In the last several decades, the functionally graded materials (FGMs) have been thoroughly investigated because of its ability to optimize mechanical behaviors by setting the material parameters into some unique function forms. So, with such attractive and practical advantages, the FGMs have been widely applied into various fields, for example the energy conversion fields, transportation, model-cutting tools, surface wrinkling, semiconductors manufacture and bio-systems [1-5].

As for the FGM vessels under various loading conditions, there has been a large number of researches exploring its elastic and elastoplastic responses by different methods [6-12]. By functionalizing the elastic modulus into an exponential form in radial direction, You. et al. [6] obtained the exact elastic expressions of stress components and radial displacement of an inner-pressurized FGM cylinder; by using a power-function-form Young's modulus and keeping the Poisson's ratio constant,

Tutuncu and Ozturk [7], analyzed the elastic behaviors of a FGM vessels with mechanical loadings; instead with a linear function form of elastic modulus, Shi et al [8] gave the elastic results of the problem same with [7].

The above works merely focus on the problems subjected to mechanical loads. Smart materials with piezomagnetic and piezoelectric effects can be often exposed to both mechanical and magnetic fields [13-22], in which the coupling mechanical and magnetic effects can candidate for tailoring intelligent structures with numerous utilization. For example, authors [23-28] have studied the effects of magnetic field on the elastic constants of magnetic elastomers and found that the material become stiffer in a magnetic field. Therefore, by this way, it is rather vital and necessary to have a better understanding of the magnetic-field's impacts on the materials mechanical behaviors, uniquely for the FGMs specializing material properties purposely by the variation of material parameters.

Recently, by assuming the elastic modulus and magnetic permeability into as power series forms (for example in [35], by setting $\mu(r) = \mu_0 r^{\beta}$ and $E(r) = E_0 r^{\beta}$) along the concerned directions, the analytical results both of displacements and the components of stresses and strains have been derived in [29-36] correspondingly when dealing with the inner pressurized FGM hollow cylinder problems. For the same problem, some other situations like the Young's Modulus represented with an exponential function form while the magnetic permeability assumed constant have also thoroughly researched [37-40]. However, all these works ignored the influences from Poisson's ratio which has been reflected in [41-43] by analyzing the radical displacements and stress components of the mechanical loaded FGM tubes under a uniform magnetic field. Furthermore, the FGMs can be literally and theoretically considered as being composed by grading its different components in certain directions by some specific volume fraction forms; but if the expressions of the components' volume fractions are set with inappropriate forms, rather complex mathematical derivation efforts and complicated results originate [44]. Acknowledging this fact, in this paper we define the volume fraction of each phase of the FGMs in an exponential form with three variables (same with [42, 45]) which can keep consistent with the patterns presented in lots of pre-researches merely by adjusting these three indexes.

For this work, we have studied the inner pressurized FGM tubes in uniform magnetic fields. In Section 2, the theoretical derivation works are finished with the analytical results of radical displacement, stress components and perturbation magnetic field vector. Then, the effects from Poisson ratio, magnetic intensity and the parameter *n* in volume fraction function are discussed in Section 3. Finally, Section 4 gives some conclusions.

**2 Theoretical analyses**

The configuration of an inner pressurized FGM tube under uniform magnetic fields within the cylindrical polar coordinate $(r,\theta,z)$ is displayed in Fig. 1; the stress boundary conditions are $\sigma_r|_{r=a} = -p$ and $\sigma_r|_{r=b} = 0$ (*a*, *b*-the inner and outer radii). In this paper, the FGM tube consists of two distinct materials A and B and the volume fraction of material A is assumed to vary in radical direction with an

expression as

$$c(r) = c_0 + c_1 (r/b)^n \tag{1}$$

where $c_0$, $c_1$ and $n$ are three material parameters.

By [46-48], the average stress and strain in a representative volume element $V$ can be defined as

$$\sigma = \frac{1}{V}\int_V \hat{\sigma}(x)dx, \qquad \varepsilon = \frac{1}{V}\int_V \hat{\varepsilon}(x)dx$$
$$\sigma^{(i)} = \frac{1}{V_i}\int_{V_i} \hat{\sigma}^{(i)}(x)dx, \qquad \varepsilon^{(i)} = \frac{1}{V_i}\int_{V_i} \hat{\varepsilon}^{(i)}(x)dx \tag{2}$$

where $\hat{\sigma}$ and $\hat{\varepsilon}$ the stress and strain fields over a representative volume element (RVE) and $\hat{\sigma}^{(i)}$, $\hat{\varepsilon}^{(i)}$ the constituents; $\sigma$, $\varepsilon$ the overall volume average stress of a RVE and $\sigma^{(i)}$, $\varepsilon^{(i)}$ the component $i$ with volume $V_i$.

For the case in this paper, by the Voigt method and uniform strain field assumption, the stress and strain can be respectively expressed as

$$\sigma = c(r)\sigma^{(1)} + [1 - c(r)]\sigma^{(2)} \tag{3}$$

$$\varepsilon_\theta^{(1)} = \varepsilon_\theta^{(2)} = \varepsilon_\theta, \quad \varepsilon_r^{(1)} = \varepsilon_r^{(2)} = \varepsilon_r \tag{4}$$

where $i=1, 2$ denotes material A and B; $\sigma^{(1)}$ and $\sigma^{(2)}$ are the average stress of material A and B; $\varepsilon_r^{(i)}$ and $\varepsilon_\theta^{(i)}$ represent the radial and circumferential strains.

For linear elastic deformation, we have

$$\varepsilon_r = \frac{du}{dr}, \qquad \varepsilon_\theta = \frac{u}{r} \tag{5}$$

where $u$ stands for the radical displacement. For each component of the FGM tube the Hooke's law can be written as

$$\sigma_r^{(i)} = \lambda_i \varepsilon_\theta^{(i)} + (\lambda_i + 2G_i)\varepsilon_r^{(i)}$$
$$\sigma_\theta^{(i)} = (\lambda_i + 2G_i)\varepsilon_\theta^{(i)} + \lambda_i \varepsilon_r^{(i)} \tag{6}$$
$$\sigma_z^{(i)} = \lambda_i \left(\varepsilon_\theta^{(i)} + \varepsilon_r^{(i)}\right)$$

where $\lambda_i$ and $G_i$ the Lamé constants; $\sigma_r^{(i)}$, $\sigma_\theta^{(i)}$ and $\sigma_z^{(i)}$ the stress components in radial, circumferential and axial directions, respectively.

Substituting equation (6) into equation (3) gives the average stress components of the FGM tube as

$$\sigma_r = \bar{\lambda}\frac{u}{r} + (\bar{\lambda} + 2\bar{G})\frac{du}{dr}$$
$$\sigma_\theta = (\bar{\lambda} + 2\bar{G})\frac{u}{r} + \bar{\lambda}\frac{du}{dr} \tag{7}$$
$$\sigma_z = \bar{\lambda}\left(\frac{u}{r} + \frac{du}{dr}\right)$$

where

$$\bar{\lambda} = c(r)\lambda_1 + [1-c(r)]\lambda_2$$
$$\bar{G} = c(r)G_1 + [1-c(r)]G_2 \tag{8}$$

With the following assumptions: (a) each material components of the FGM tube is non-ferromagnetic and non-ferroelectric; (b) the Thompson effects are omittable; (c) the displacement electric currents are ignored, then for elastic medium with perfect conductions the simplified electrodynamics Maxwell's equations can be written as [35, 36]

$$\vec{J} = \nabla \times \vec{h}, \quad \vec{h} = \nabla \times (\vec{U} \times \vec{H}), \quad \text{div }\vec{h}=0,$$
$$\vec{e} = -\mu(r)\left(\frac{\partial \vec{U}}{\partial t} \times \vec{H}\right), \quad \nabla \times \vec{e} = -\mu(r)\frac{\partial \vec{h}}{\partial t} \tag{9}$$

where $\vec{J}$, $\vec{h}$, $\vec{U}$, $\vec{H}$, $\vec{e}$ and $t$ are correspondingly the electric current density, perturbation of magnetic field, electrical displacement, magnetic intensity, perturbation of electric field vector and time variable.

Tanking an initial magnetic field vector $\vec{H}(0,0,H_z)$ into equation (9) yields

$$\vec{U} = (u,0,0), \quad \vec{e} = \mu(r)\left(0, H_z \frac{\partial u}{\partial t}, 0\right), \quad \vec{h} = (0,0,h_z),$$
$$\vec{J} = \left(0, -\frac{\partial h_z}{\partial r}, 0\right), \quad h_z = -H_z\left(\frac{\partial u}{\partial r} + \frac{u}{r}\right) \tag{10}$$

Then, by $\vec{f} = \mu(r)(\vec{J} \times \vec{H})$ the radical Lorentz's stress $f_r$ can be induced as

$$f_r = H_z^2 \mu(r)\left(\frac{\partial^2 u}{\partial r^2} + \frac{1}{r}\frac{\partial u}{\partial r} - \frac{u}{r^2}\right) \tag{11}$$

where $\mu(r)$, the magnetic permeability of the FGM tube, can be expressed by the Voigt method as

$$\mu(r) = c(r)\mu_1 + [1 - c(r)]\mu_2 \tag{12}$$

Substituting equations (7), (11) and (12) into equilibrium equation

$$\frac{d\sigma_r}{dr} + \frac{\sigma_r - \sigma_\theta}{r} + f_r = 0 \tag{13}$$

obtains the governing ordinary differential equation for the radial displacement $u$

$$r(\phi_1 - \phi_2 r^n)\frac{d^2 u}{dr^2} + (\phi_1 - \phi_3 r^n)\frac{du}{dr} - (\phi_1 + \phi_4 r^n)\frac{u}{r} = 0 \tag{14}$$

where

$$\phi_1 = c_0\left(\lambda_1 + 2G_1 + \mu_1 H_z^2\right) + (1-c_0)\left(\lambda_2 + 2G_2 + \mu_2 H_z^2\right)$$
$$\phi_2 = c_1\left(\lambda_2 + 2G_2 + \mu_2 H_z^2 - \lambda_1 - 2G_1 - \mu_1 H_z^2\right)/b^n$$
$$\phi_3 = (n+1)\phi_2 + c_1 n H_z^2(\mu_1 - \mu_2)/b^n$$
$$\phi_4 = n(\lambda_2 - \lambda_1)c_1/b^n - \phi_2$$

(15)

## 2.1 Case 1: $\phi_1 \neq 0$

For the case $\phi_1 \neq 0$, equation (14) can be rearranged as

$$r\left(1 - \frac{\phi_2}{\phi_1}r^n\right)\frac{d^2u}{dr^2} + \left(1 - \frac{\phi_3}{\phi_1}r^n\right)\frac{du}{dr} - \left(1 + \frac{\phi_4}{\phi_1}r^n\right)\frac{u}{r} = 0 \quad (16)$$

For convenience, by setting $x = \chi(r) = \frac{\phi_2}{\phi_1}r^n$, equation (16) can be rewritten as

$$x^2(1-x)\frac{d^2u}{dx^2} + x\left(1 - \frac{n-1+\phi_3/\phi_2}{n}x\right)\frac{du}{dx} - \frac{1}{n^2}\left(1 + \frac{\phi_4}{\phi_2}x\right)u = 0 \quad (17)$$

According to [49], equation (17) can be solved as

$$u(r) = C_1 r F(\alpha, \beta, \delta; x) + C_2 \frac{1}{r} F(\alpha - \delta + 1, \beta - \delta + 1, 2 - \delta; x) \quad (18)$$

where $C_1$ and $C_2$ are constants; $F$ is the hypergeometric function defined in $|x| < 1$ with a power series form as

$$F(\alpha, \beta, \delta; x) = 1 + \sum_{m=1}^{\infty} \frac{C_{\alpha+m-1}^m C_{\beta+m-1}^m}{C_{\delta+m-1}^m} x^m \quad (19)$$

in where

$$\delta = 1 + \frac{2}{n}, \quad \alpha = \frac{\sqrt{(\phi_3/\phi_2 - 1)^2 - 4\phi_4/\phi_2} + \phi_3/\phi_2 + 1}{2n}, \quad \beta = \frac{\phi_3/\phi_2 + 1}{n} - \alpha \quad (20)$$

Note that equation (18) holds throughout by the following.

Rearrange the radical displacement

$$u(r) = C_1 P(r) + C_2 Q(r) \quad (21)$$

where the specific form of $P(r)$ and $Q(r)$ and their derivatives respect to $r$ are

$$\begin{aligned}
P(r) &= rF(\alpha,\beta,\delta;x) \\
Q(r) &= \frac{1}{r}F(\alpha-\delta+1,\beta-\delta+1,2-\delta;x) \\
\frac{dP(r)}{dr} &= \frac{n\alpha\beta x}{\delta}F(\alpha+1,\beta+1,\delta+1;x)+\frac{P(r)}{r} \\
\frac{dQ(r)}{dr} &= \frac{1}{r}\left[\frac{n(\alpha-\delta+1)(\beta-\delta+1)x}{(2-\delta)r}F(\alpha-\delta+2,\beta-\delta+2,3-\delta;x)-Q(r)\right]
\end{aligned} \quad (22)$$

Then, by equation (7) and equation (10)$_5$ the stress components and the perturbation of magnetic field can be derived as

$$\begin{aligned}
\sigma_r &= (\bar{\lambda}+2\bar{G})\left[C_1\frac{dP(r)}{dr}+C_2\frac{dQ(r)}{dr}\right]+\bar{\lambda}\left[C_1\frac{P(r)}{r}+C_2\frac{Q(r)}{r}\right] \\
\sigma_\theta &= \bar{\lambda}\left[C_1\frac{dP(r)}{dr}+C_2\frac{dQ(r)}{dr}\right]+(\bar{\lambda}+2\bar{G})\left[C_1\frac{P(r)}{r}+C_2\frac{Q(r)}{r}\right] \\
\sigma_z &= \bar{\lambda}\left[C_1\frac{dP(r)}{dr}+C_2\frac{dQ(r)}{dr}+C_1\frac{P(r)}{r}+C_2\frac{Q(r)}{r}\right] \\
h_z &= -H_z\left[C_1\frac{dP(r)}{dr}+C_2\frac{dQ(r)}{dr}+C_1\frac{P(r)}{r}+C_2\frac{Q(r)}{r}\right]
\end{aligned} \quad (23)$$

By natural boundary conditions $\sigma_r|_{r=a}=-p$ and $\sigma_r|_{r=b}=0$, there is

$$\begin{aligned}
C_1 &= -p\left[(\bar{\lambda}(b)+2\bar{G}(b))Q'(b)+\bar{\lambda}(b)Q(b)/b\right]/C_0 \\
C_2 &= p\left[(\bar{\lambda}(b)+2\bar{G}(b))P'(b)+\bar{\lambda}(b)P(b)/b\right]/C_0
\end{aligned} \quad (24)$$

where

$$\begin{aligned}
C_0 &= \left[(\bar{\lambda}(b)+2\bar{G}(b))Q'(b)+\bar{\lambda}(b)Q(b)/b\right]\left[(\bar{\lambda}(a)+2\bar{G}(a))P'(a)+\bar{\lambda}(a)P(a)/a\right] \\
&\quad -\left[(\bar{\lambda}(a)+2\bar{G}(a))Q'(a)+\bar{\lambda}(a)Q(a)/a\right]\left[(\bar{\lambda}(b)+2\bar{G}(b))P'(b)+\bar{\lambda}(b)P(b)/b\right]
\end{aligned} \quad (25)$$

In the following, some special situations are discussed.

(a) If $c_0 \neq 0$ and $c_1=0$ which indicates that the tube contains only one material with graded parameters along radical direction, then equation (14) reduces to the well-known Eulerian equation

$$r^2\frac{d^2u}{dr^2}+r\frac{du}{dr}-u=0 \tag{26}$$

with solution as

$$u(r)=\frac{pa^2}{2(b^2-a^2)}\left\{\frac{r}{c_0(\lambda_1+G_1)+(1-c_0)(\lambda_2+G_2)}+\frac{b^2}{r[c_0G_1+(1-c_0)G_2]}\right\} \tag{27}$$

By which equation (23) can be rewritten as

$$\begin{aligned}\sigma_r&=\frac{pa^2}{b^2-a^2}(1-\frac{b^2}{r^2})\\ \sigma_\theta&=\frac{pa^2}{b^2-a^2}(1+\frac{b^2}{r^2})\\ \sigma_z&=\frac{pa^2[c_0\lambda_1+(1-c_0)\lambda_2]}{(b^2-a^2)[c_0(\lambda_1+G_1)+(1-c_0)(\lambda_2+G_2)]}\\ h_z&=-\frac{pa^2H_z}{(b^2-a^2)[c_0(\lambda_1+G_1)+(1-c_0)(\lambda_2+G_2)]}\end{aligned} \tag{28}$$

(b) If $n=0$, the tube becomes isotropic and equations (27) and (28) change into the results given by [50].

It should be noted for the above two cases that the Lorentz's stress in the radical direction approaches to zero which can be verified by taking equation (26) into equation (11).

(c) If $H_z=0$ and $c_1=-c_0k$, equations (21) and (23) reduces to equations (12) and (13) gotten by [42].

## 2.2 Case 2: $\phi_1=0$

In this case, equation (14) can be simplified as

$$\phi_2r^2\frac{d^2u}{dr^2}+\phi_3r\frac{du}{dr}+\phi_4u=0 \tag{29}$$

with solution

$$u(r) = C_3 r^{m_1} + C_4 r^{m_2} \tag{30}$$

Then, by the above results we derive equation (23) as

$$\begin{aligned}
\sigma_r &= C_3 r^{m_1-1}\left[(m_1+1)\bar{\lambda} + 2m_1\bar{G}\right] + C_4 r^{m_2-1}\left[(m_2+1)\bar{\lambda} + 2m_2\bar{G}\right] \\
\sigma_\theta &= C_3 r^{m_1-1}\left[(m_1+1)\bar{\lambda} + 2\bar{G}\right] + C_4 r^{m_2-1}\left[(m_2+1)\bar{\lambda} + 2\bar{G}\right] \\
\sigma_z &= \bar{\lambda}\left[C_3(m_1+1)r^{m_1-1} + C_4(m_2+1)r^{m_2-1}\right] \\
h_z &= -H_z\left[C_3(m_1+1)r^{m_1-1} + C_4(m_2+1)r^{m_2-1}\right]
\end{aligned} \tag{31}$$

where $m_1 = \dfrac{1}{2} - \dfrac{\phi_3}{2\phi_2} + \dfrac{1}{2}\sqrt{\left(\dfrac{\phi_3}{\phi_2}-1\right)^2 - \dfrac{4\phi_4}{\phi_2}}$ and $m_2 = \dfrac{1}{2} - \dfrac{\phi_3}{2\phi_2} - \dfrac{1}{2}\sqrt{\left(\dfrac{\phi_3}{\phi_2}-1\right)^2 - \dfrac{4\phi_4}{\phi_2}}$ and $C_3$

and $C_4$ given by the boundary condition $\sigma_r|_{r=a} = -p$ and $\sigma_r|_{r=b} = 0$ as

$$\begin{aligned}
C_3 &= -pb^{m_2-1}\left[(m_2+1)\bar{\lambda}(b) + 2m_2\bar{G}(b)\right]/\bar{C}_0 \\
C_4 &= pb^{m_1-1}\left[(m_1+1)\bar{\lambda}(b) + 2m_1\bar{G}(b)\right]/\bar{C}_0
\end{aligned} \tag{32}$$

where

$$\begin{aligned}
\bar{C}_0 &= a^{m_1-1}b^{m_2-1}\left[(m_1+1)\bar{\lambda}(a) + 2m_1\bar{G}(a)\right]\left[(m_2+1)\bar{\lambda}(b) + 2m_2\bar{G}(b)\right] \\
&\quad - a^{m_2-1}b^{m_1-1}\left[(m_1+1)\bar{\lambda}(b) + 2m_1\bar{G}(b)\right]\left[(m_2+1)\bar{\lambda}(a) + 2m_2\bar{G}(a)\right]
\end{aligned} \tag{33}$$

Therefore, we have

(a) When material A and material B have same Poisson's ratio and with conditions as $E_2 - E_1 = E_0, E_2/E_0 = c_0, m_2 - m_1 = m_0, m_2/m_0 = c_0, c_1 = -1$, equation (29) reduces to the equation (4) in [36], i.e.,

$$r^2 \frac{d^2u}{dr^2} + (1+\bar{\zeta}n)r\frac{du}{dr} + (\bar{\eta}n - 1)u = 0 \tag{34}$$

where

$$\bar{\zeta} = \frac{(E_2 - E_1)(1-v)}{(E_2 - E_1)(1-v) + H_z^2(\mu_2 - \mu_1)(1+v)(1-2v)}$$
$$\bar{\eta} = \frac{(E_2 - E_1)v}{(E_2 - E_1)(1-v) + H_z^2(\mu_2 - \mu_1)(1+v)(1-2v)} \quad (35)$$

(b) When $H_z=0$, equation (34) takes a similar form with [7], i.e.,

$$r^2 \frac{d^2u}{dr^2} + (1+n)r\frac{du}{dr} + \left(\frac{nv}{1-v} - 1\right)u = 0 \quad (36)$$

Then, same solutions for displacement and stress components can be determined.

## 3 Numerical examples and discussions

In the use of the following dimensionless quantities $\bar{r} = r/b$, $\bar{a} = a/b$, $\bar{\sigma}_{ij} = \sigma_{ij}(r)/p$, $\bar{u} = u(r)E_2/(bp)$ and $\bar{h}_z = h_z(r)E_2/(H_z p)$ in where $a$ and $b$ are the inner and outer radii; $p$ the internal pressure; $Hz$ the magnetic intensity; $E_2$ material B's elastic modulus, this numerical part provides some examples to explore the influences from Poisson's ration, magnetic density and index $n$. And the selected configuration and material parameters have listed in Table 1.

### 3.1 Effects of Poisson's ratio

In this part, the effects of Poisson's ratio on FGM tube in uniform magnetic fields have been discussed. It assumes that the FGM tube is made by two different materials, three sets of Poisson's rations are chosen, i.e., $v_1 = 0.3$, $v_2 = 0.2$; $v_1 = v_2 = 0.3$; $v_1 = 0.3$, $v_2 = 0.4$.

Indicated from Fig. 2, it can be concluded that the Poisson's ration has an obvious influence on the distribution of radical displacement for the FGM tube under uniform magnetic field when compared with the results of [42] without magnetic fields. Aside from the displacement, the Poisson's ratio displays similar effects on stress components revealing as Figs. 3-5; specifically, the trend of the distribution of radical stress increases with the decrease of Poisson's ratio contrary to the circumferential stress (see Figs. 3 and 4); as for the axial stress, the influence from Poisson's ratio is more pronounced closer to the inner tube's surface meanwhile the minimum value achieves in the outer surface as exhibited in Fig. 5.

From Fig. 6, it can be seen that the magnetic field vector distributes almost horizontally for each set of chosen Poisson's ratio while all the values are negative.

**3.2 Effects of magnetic intensity**

Based on equations (21), (23), (30) and (31), this part investigates the effects of magnetic intensity on the mechanical responses of the FGMs tube and makes a comparison with the situation ignoring the magnetic field. All the results have been graphed in Figs. 7-9.

As illustrated by Fig. 7, the radical displacement, both of the cases with/without magnetic field, declines from the inner surface to the outer surface while the differences are verified by the larger radical displacement when $H_z = 0$ than that when $H_z = 2.23 \times 10^9 \ A/m$. Fig. 8 shows evident distribution trends among the axial, radical and circumferential stresses: while the value of radical stress increases with the increment of radius, the other two just decrease; for different magnetic field density, the radical stress keep nearly unchanged but certain differences happen to the stresses in axial and circumferential direction. From Fig. 9, it can be acknowledged that the magnetic field affects the perturbation of magnetic vector significantly. For example, when there is no magnetic field the perturbation of magnetic vector is equal to zero but weakly increases along the radical direction when magnetic field being with a certain intensity.

**3.3 Effects of the parameter *n***

Since the volume fractions of materials A and B are determined by the parameter *n* as shown in equation (1), it seems rather necessary to detect its impacts on the performance of the FGM tube (see Figs. 10-14). In this section, the parameters are selected as *n*=1.5, 3 and 5.

Fig. 10 shows that the radical displacement declines with the increment of parameter *n* and reaches to its minimum value at outer surface. For the distribution of stress components with natural boundary conditions $\bar{\sigma}_r|_{\bar{r}=\bar{a}}=0$ and $\bar{\sigma}_r|_{\bar{r}=\bar{b}}=0$, parameter *n* just inconsiderably influences the distribution of the radical stress, i.e., the differences mainly focusing on the middle part of the tube with a less than 3% accretion from *n*=1.5 to *n*=3 while a less than 3.5% increment from *n*=3 to *n*=5 (see Fig. 11); for the stresses in circumferential and axial direction, the maximum difference (more than 50%) is located at the outer surface when parameter *n* changing from 3 to 5 (see Fig. 12 and Fig. 13). But for the perturbation of magnetic field vector, the differences between two parameters reduce with the augment of parameter *n* while with a flat distribution along the radical direction (see Fig. 14). Moreover, as shown in Fig. 14, the values of the magnetic field for different *n* are negative and incline to zero with the growth of *n*.

## 4 Conclusions

In summary, this paper studied the mechanical response of an inner pressurized FGM cylinder composed by two different materials within uniform magnetic fields. By assuming the volume fraction of material component into an exponential function form, we derived the analytical expressions of the radical displacement, the stress components and the perturbation magnetic vector by the Voigt method. Furthermore, the effects of the Poisson's ratio, the magnetic intensity and the parameter *n* are discussed. From the results of the numerical part it can be concluded that the both the Poisson's ratio and the parameter *n* have a clear influence on the radical displacement, the axial stress and the perturbation of the magnetic field vector. And by comparing with the results with our previous work, essential differences of the mechanical responses of the FGM tube has been illustrated between the situation with and without the action of magnetic field. These results obtained in this paper can serve as important contribution to the design and development of the FGM structures within multi-physical fields.

**Compliance with Ethical Standards**

**Funding** This study was funded by National Natural Science Foundation of China (No. 11772041).

**Conflict of Interest** The authors declare that they have no conflict of interest.

**Tables**

**Table 1** The selected configuration and material parameters

| $\bar{a}$ | $c_0$ | $c_1$ | $E_1$ | $E_2$ | $\mu$ |
|---|---|---|---|---|---|
| 0.7 | 1 | 1 | 210GPa | 70GPa | $4\pi \times 10^{-7} \, H/m$ |

**Figure captions**

**Fig. 1** Schematic diagram of a long FGM tube subjected to internal pressure in a uniform magnetic field $H_z$.

**Fig. 2** Evolution of the radial displacement with different Poisson's ratio ($n$=1.5, $H_z$=2.23×10$^9$ A/m).

**Fig. 3** Evolution of the radial stress with different Poisson's ratio ($n$=1.5, $H_z$=2.23×10$^9$ A/m).

**Fig. 4** Evolution of the circumferential stress with different Poisson's ratio ($n$=1.5, $H_z$=2.23×10$^9$ A/m).

**Fig. 5** Evolution of the axial stress with different Poisson's ratio ($n$=1.5, $H_z$=2.23×10$^9$ A/m).

**Fig. 6** Evolution of the perturbation of magnetic field vector with different Poisson's ratio ($n$=1.5, $H_z$=2.23×10$^9$ A/m).

**Fig. 7** Comparison of two different magnetic intensity vectors ($H_z$=2.23×10$^9$A/m in this work and $H_z$=0 in Ref. [42]) for the radical displacement ($n$=1.5, $v_1$=0.2, $v_2$=0.3).

**Fig. 8** Comparison of two different magnetic intensity vectors ($H_z$=2.23×10$^9$A/m in this work and $H_z$=0 in Ref. [42]) for the stresses ($n$=1.5, $v_1$=0.2, $v_2$=0.3).

**Fig. 9** Comparison of two different magnetic intensity vectors ($H_z$=2.23×10$^9$A/m in this work and $H_z$=0 in Ref. [42]) for the perturbation of magnetic field vector ($n$=1.5, $v_1$=0.2, $v_2$=0.3).

**Fig. 10** Evolution of the radical displacement with different parameter $n$ ($v_1$=0.2, $v_2$=0.3, $H_z$=2.23×10$^9$ A/m).

**Fig. 11** Evolution of the radical stress with different parameter $n$ ($v_1$=0.2, $v_2$=0.3, $H_z$=2.23×10$^9$ A/m).

**Fig. 12** Evolution of the circumferential stress with different parameter $n$ ($v_1$=0.2, $v_2$=0.3, $H_z$=2.23×10$^9$ A/m).

**Fig. 13** Evolution of the axial stress with different parameter $n$ ($v_1$=0.2, $v_2$=0.3, $H_z$=2.23×10$^9$ A/m).

**Fig. 14** Evolution of the perturbation of magnetic field vector with different parameter $n$ ($v_1$=0.2, $v_2$=0.3, $H_z$=2.23×10$^9$ A/m).

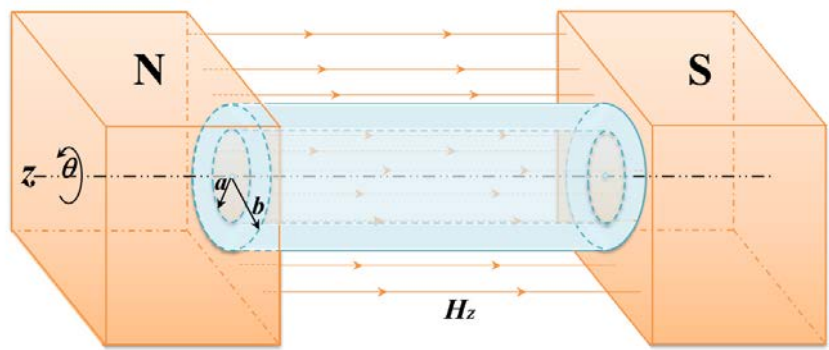

**Fig. 1** Schematic diagram of a long FGM tube subjected to internal pressure in a uniform magnetic field $H_z$.

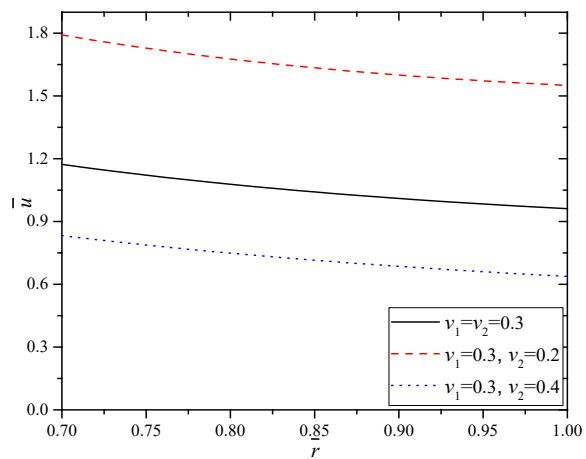

**Fig. 2** Evolution of the radial displacement with different Poisson's ratio ($n$=1.5, $H_z$=2.23×10$^9$ A/m).

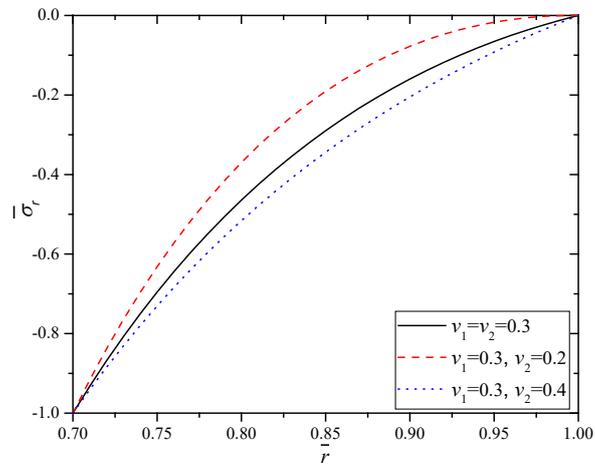

**Fig. 3** Evolution of the radial stress with different Poisson's ratio ($n$=1.5, $H_z$=2.23×10$^9$ A/m).

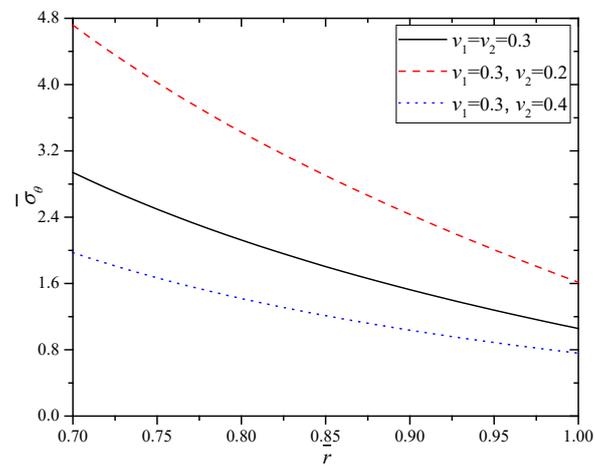

**Fig. 4** Evolution of the circumferential stress with different Poisson's ratio ($n$=1.5, $H_z$=2.23×10$^9$ A/m).

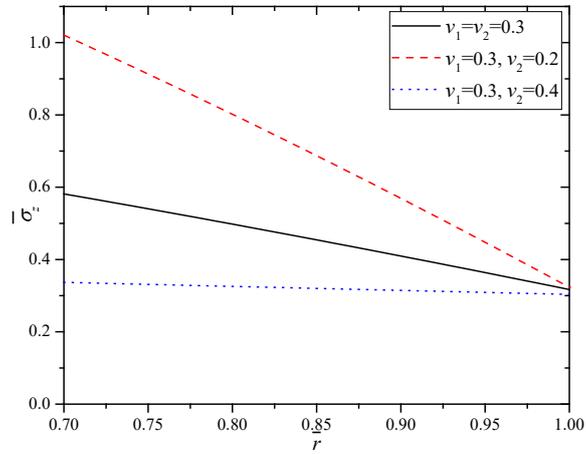

**Fig. 5** Evolution of the axial stress with different Poisson's ratio ($n=1.5$, $H_z=2.23\times10^9$ A/m).

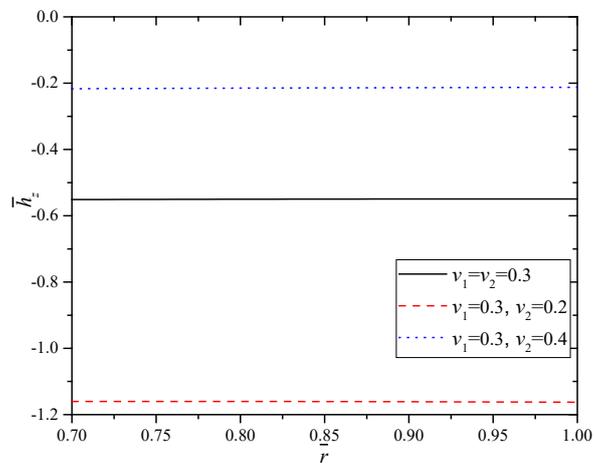

**Fig. 6** Evolution of the perturbation of magnetic field vector with different Poisson's ratio ($n=1.5$, $H_z=2.23\times10^9$ A/m).

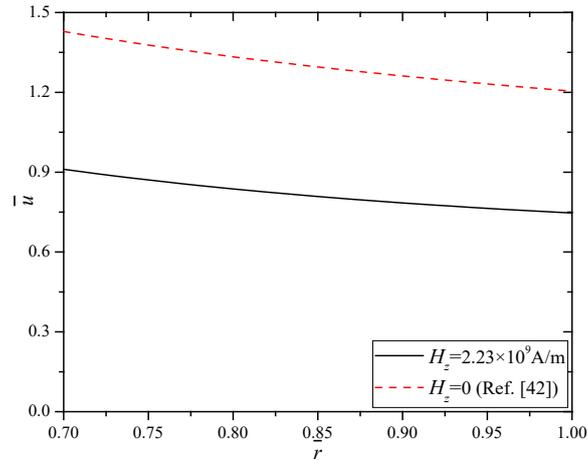

**Fig. 7** Comparison of two different magnetic intensity vectors ($H_z=2.23\times10^9$A/m in this work and $H_z=0$ in Ref. [42]) for the radical displacement ($n=1.5$, $v_1=0.2$, $v_2=0.3$).

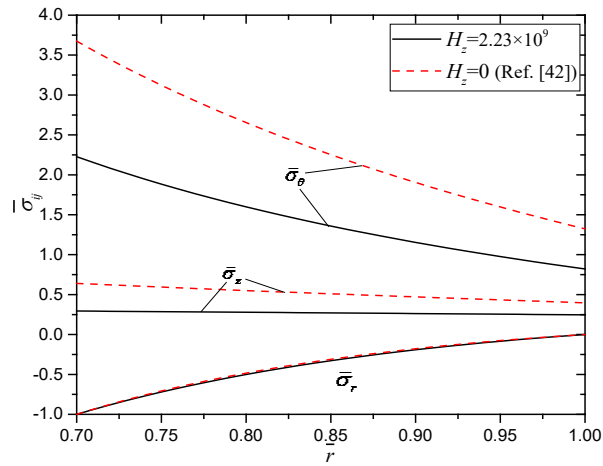

**Fig. 8** Comparison of two different magnetic intensity vectors ($H_z=2.23\times10^9$A/m in this work and $H_z=0$ in Ref. [42]) for the stresses ($n=1.5$, $v_1=0.2$, $v_2=0.3$).

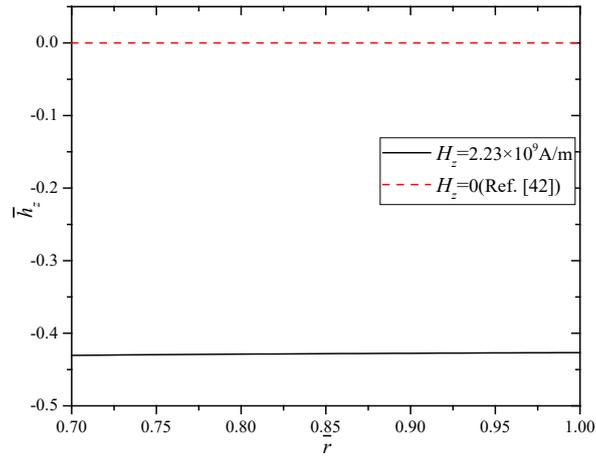

**Fig. 9** Comparison of two different magnetic intensity vectors ($H_z=2.23\times10^9$A/m in this work and $H_z=0$ in Ref. [42]) for the perturbation of magnetic field vector ($n=1.5$, $v_1=0.2$, $v_2=0.3$).

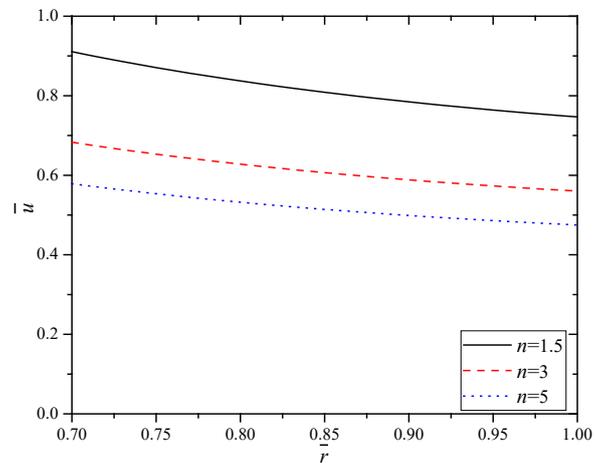

**Fig. 10** Evolution of the radical displacement with different parameter $n$ ($v_1=0.2$, $v_2=0.3$, $H_z=2.23\times10^9$ A/m).

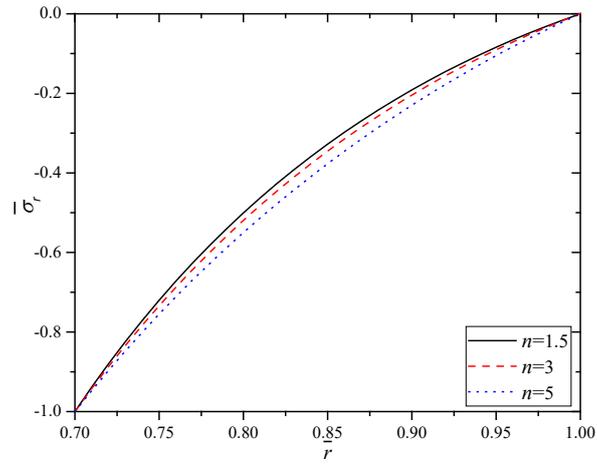

**Fig. 11** Evolution of the radical stress with different parameter $n$ ($v_1$=0.2, $v_2$=0.3, $H_z$=2.23×10$^9$ A/m).

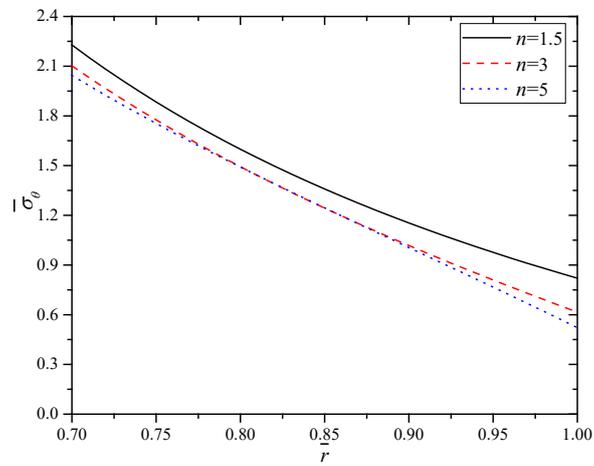

**Fig. 12** Evolution of the circumferential stress with different parameter $n$ ($v_1$=0.2, $v_2$=0.3, $H_z$=2.23×10$^9$ A/m).

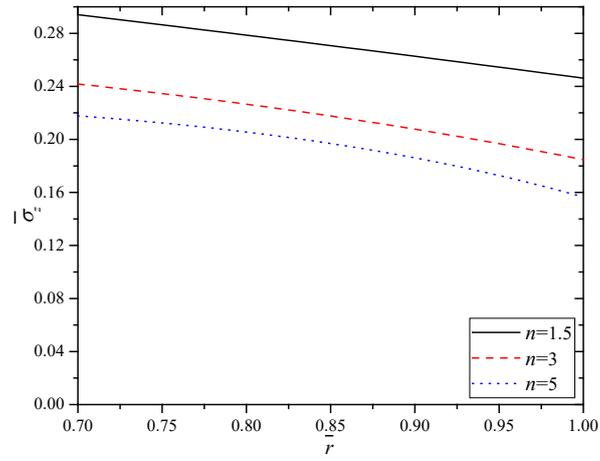

**Fig. 13** Evolution of the axial stress with different parameter $n$ ($v_1$=0.2, $v_2$=0.3, $H_z$=2.23×10$^9$ A/m).

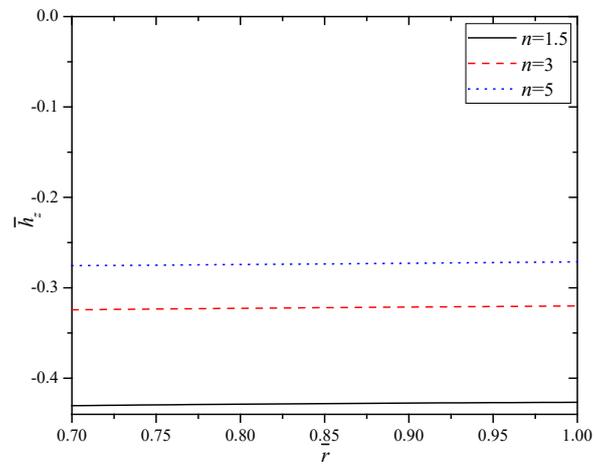

**Fig. 14** Evolution of the perturbation of magnetic field vector with different parameter $n$ ($v_1$=0.2, $v_2$=0.3, $H_z$=2.23×10$^9$ A/m).